\documentclass[twocolumn,showpacs,preprintnumbers,amsmath,amssymb]{revtex4}
 
\usepackage{graphicx}% Include figure files
\usepackage{dcolumn}% Align table columns on decimal point
\usepackage{bm}% bold math

\begin{document}
 
\preprint{Physical Review Letters. In press, 2005}

\title{Auto and crosscorrelograms for the spike response of LIF neurons
with slow synapses}

\author{ {\bf\sc Rub\'en Moreno-Bote (1) and N\'estor Parga (2,3)}}
\affiliation{\it 
(1) Center for Neural Science,
New York University,
New York, NY 10003-6621, USA. \\
\it(2) Dept. de F\'{\i}sica Te\'orica,
 Universidad Aut\'onoma de Madrid,
Cantoblanco, 28049 Madrid, Spain.\\
\it (3) Center for Theoretical Neuroscience, Center for Neurobiology and
Behavior, Columbia University, NY 10032-2695, USA.
}

\def    \be            {\begin{equation}}
\def    \ee            {\end{equation}}
\def    \bea           {\begin{eqnarray}}
\def    \eea           {\end{eqnarray}}
\def    \nn            {\nonumber}
\def    \raw           {\rightarrow}
\def    \law           {\leftarrow}
 
\newcommand{\bi}{\begin{itemize}}
\newcommand{\ei}{\end{itemize}}

\begin{abstract}                                                               
An analytical description of the response properties of simple but
realistic neuron models in the presence of noise is still lacking. We
determine completely up to the second order the firing statistics of a
single and a pair of leaky integrate-and-fire neurons (LIFs)
receiving some common slowly filtered white noise. In particular, 
the auto- and cross-correlation functions of the output
spike trains of pairs of cells are obtained from an improvement of
the adiabatic approximation introduced in \cite{Mor+04}.
These two functions define the firing
variability and firing synchronization between neurons, and are of
much importance for understanding neuron communication.
\end{abstract}


\maketitle

%%%%%%%%%%%%%%%%%%% INTRODUCTION %%%%%%%%%%%%

The variability of the spike trains of cortical neurons and their
correlations might constraint the coding capabilities of the brain
\cite{Sha+98otros}, but they can also reflect the strategies the
brain uses to decipher the stimuli arriving from the world
\cite{Sin99Rotros}. Neurons in cortex fire with high
variability resembling Poisson spike trains \cite{Sof+93}, and nearby
pairs of cortical neurons fire in a correlated fashion
\cite{Sha+98otros}, reflecting the presence of some common source of
noise. These variability and correlation of the spike trains affect
the firing statistics of a neuron receiving those inputs
\cite{Sal+00,Mor+02}. It has been shown that the large variability 
observed {\em in vivo} can be accounted for by neuron models operating
in a regime in which the membrane time constant, $\tau_m$, becomes
shorter or comparable to the synaptic decay constants, $\tau_s$, due
to spontaneous background activity ($\tau_s \ge
\tau_m$) \cite{Mor+05,Svi-Des}. However, very little 
progress has been made in providing analytical tools to describe
such variability and correlations found in cortex.

In this Letter we study analytically the variability and correlations
in the firing responses of pairs of LIF neurons receiving both common
and independent sources of white noise input filtered by synapses in
the regime $\tau_s \ge \tau_m$. For a single neuron we obtain the
firing rate, the autocorrelation function of its output spike train
(ACF), the Fano factor of the spike count, $F_N$. For a pair of
cells, we obtain the crosscorrelation function of their output spike
trains (CCF) and the correlation coefficient of their spike counts,
$\rho$.  These results characterize completely the firing response of
these spiking neurons up to second order, and open the possibility for
a principled way of including synchrony effects 
in the modeling of biologically plausible spiking neural
networks.

%%%%%%%%%%%%%%%%%%%%%%%% MODEL     %%%%%%%%%%%%%%%%%%%%

\noindent
{\bf The neuron and input models.} The membrane potential $V(t)$ of a
single LIF neuron with membrane time constant $\tau_m$ and receiving
an afferent current $I(t)$ obeys

\begin{equation}
    \tau_m \; \dot{V}= -V + \tau_m \;I(t)  \; .
    \label{eq:V}
\end{equation}

\noindent
A spike is generated when $V(t)$ reaches a threshold $\Theta$, after
which the neuron is reset to $H$, from where it continues integrating
the current \cite{Ric77}. The external input is modeled by a white
noise with mean $\mu$ and variance $\sigma^2$ \cite{Ric77} which is
filtered by synapses with decay time constant $\tau_s$, resulting in a
current described by

\begin{equation}
   \tau_s \dot{I}(t)= -I(t) + \mu + \sigma \eta(t) \; , 
   \label{eq:I}
\end{equation}

\noindent
where $\eta(t)$ is a Gaussian white noise with zero mean and unit
variance. We simplify eqs. (\ref{eq:V}-\ref{eq:I}) by performing the
linear transformations $I=\mu + z \; \sigma / \sqrt{2 \tau_s}$ and
$V=\mu \tau_m + x \;\sigma \sqrt{\tau_m / 2} $, obtaining 

\begin{eqnarray}
     \dot{x} &=& \frac{1}{\tau_m} (-x + \gamma z ) \;
\label{eq:x}
\\
    \dot{z} &=& -\frac{z}{\tau_s} + \sqrt{\frac{2}{\tau_s}} \eta(t) \; ,
\label{eq:z}
\end{eqnarray}

\noindent
with $\gamma=\sqrt{\tau_m/\tau_s}$. In the normalized potential, $x$,
the threshold and reset read $\hat \Theta = \sqrt{2}(\Theta - \mu
\tau_m)/ \sigma \sqrt{\tau_m}$ and $\hat H =\sqrt{2}(H - \mu \tau_m)/
\sigma \sqrt{\tau_m}$.

\noindent
{\bf The autocorrelation function.}  To determine the ACF, first we
describe the time evolution of the probability density of having the
neuron in the state $(x,z)$ at time $t$ given that initially the
neuron has just fired ($x=\hat H$) and $z=z_0$. The Fokker-Planck
equation (FPE) for this density, $P(x,z,t|\hat H,z_0)$, is
\cite{Mor+04}

\begin{equation}
\tau_m \frac{\partial}{\partial t} P= 
   \left[ \frac{\partial}{\partial x} (x- \gamma z) + 
    \epsilon^2  L_z \right] P + \tau_m J(z,t|z_0) \delta(x-\hat H) \; ,
   \label{eq:FPE}
\end{equation}

\noindent
where $\epsilon=\gamma=\sqrt{\tau_m/\tau_s}$ and $L_z=\frac{\partial
}{\partial z} z + \frac{\partial^{2} }{\partial^{2} z}$. $J(z,t|z_0)$
is the probability density of having a spike at time $t$ along with a
fluctuation $z$ given that $z=z_0$ at time $t=0$. This probability is
expressed as a function of the density $P$ as \cite{Mor+04}

\begin{equation}
   J(z,t|z_0)= \frac{1}{\tau_m} ( -\hat \Theta + \gamma z ) \; P(\hat
   \Theta,z,t|\hat H,z_0) \; .  \label{eq:J}
\end{equation}

\noindent
Solving the FPE (\ref{eq:FPE}) with $J(z,t|z_0)$ as a source term at
$x=\hat H$ means that each time a spike is produced, the normalized
potential $x$ is reset to $\hat H$ while $z$ keeps its same value.

The integral $\int dz J(z,t|z_0)$ expresses the
probability of having a spike at time $t$ conditioned to the fact that $z=z_0$
at time $t=0$. We define the ACF, $C(t)$,
as the probability density of firing a spike at time $t > 0$ conditioned
to the fact that at time $t=0$ there was a spike. Therefore, $C(t)$ is
the average of $\int dz J(z,t|z_0)$ with the distribution of $z_0$
conditioned to the production of a spike at time $t=0$,
$B(z_0)$. Since $B(z)$ is the distribution of $z$ at the {\em moment
of a spike}, then $B(z)=J(z) / \nu$, where $J(z)$ is the limit $t
\rightarrow \infty$ of $J(z,t|z_0)$, and $\nu$ is its normalizing
factor ($\nu=\int dz J(z)$) and also the firing rate of the LIF neuron
defined by eqs. (\ref{eq:x}-\ref{eq:z}). Therefore, the ACF is
computed as

\begin{equation}
   C(t)  =  \int dz_0 \; \frac{J(z_0)}{\nu} \;\int dz \; J(z,t|z_0) \;.
   \label{eq:C}
\end{equation}

\noindent
The solution of the FPE (\ref{eq:FPE}) and eq. (\ref{eq:C}) is
simplified by noticing that $z$ is a pure Ornstein-Uhlenbeck process,
eq. (\ref{eq:z}), and therefore its marginal distribution,
$P(z,t|z_0)$, is (see, e.g., \cite{Ric77})

\begin{equation}
   P(z,t|z_0)=  \frac{1}{\sqrt{2 \pi (1-e^{-2t/\tau_s})}} \;
              e^{-\frac{(z-z_0 \; e^{-t/\tau_s})^2}{2(1-e^{-2t/\tau_s})}} \:,
   \label{eq:FPE-z-sol}
\end{equation}

\noindent
which broadens over time and for $t \gg \tau_s$ approaches a normal
distribution, $p(z)=e^{-z^2/2}/\sqrt{2 \pi}$.

%%%%%%  RESULTS  %%%%%%%%%

\begin{figure}
\includegraphics[width=8cm,height=5cm,angle=0]{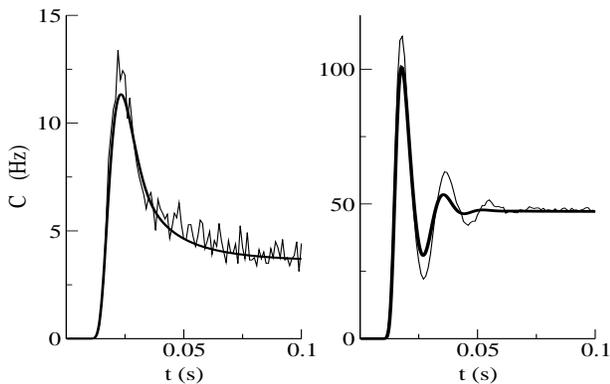}
\caption{\label{fig:grafica1}
ACF of a LIF neuron in the sub (left) and
suprathreshold regimes (right). The figures show the typical shape in
both regimes: no oscillations and a large peak in the subthreshold
regime ($\mu \tau_m <\Theta$) and damped oscillations in the
suprathreshold regime ($\mu \tau_m >\Theta$). Thick lines are the
analytical results obtained from eq.(\ref{eq:C_0}) (the sum has been
cut at $n=200$ with $t=200ms$) and thin lines correspond to the
numerical simulations of the same LIF neuron. Parameters for the
subthreshold (suprathreshold)neuron are $\mu=85Hz(115Hz)$ $\sigma^2=6Hz(3Hz)$. Other
parameters are $H=0$, $\Theta=1$, $\tau_m=10ms$ and $\tau_s=20ms$.  }
\end{figure}

\noindent
{\bf The analytical solution.} We expand both $P(x,z,t|\hat H,z_0)$ and
$J(z,t|z_0)$ in powers of $\epsilon^2$, as $P=P_0+\epsilon^2 P_1 +
0(\epsilon^4)$ and $J=J_0+\epsilon^2 J_1 + 0(\epsilon^4)$, following a
technique introduced in \cite{Mor+04} for the stationary FPE. In this
expansion, the parameter $\gamma$ in eqs (\ref{eq:FPE},\ref{eq:J}) is
assumed to be fixed. Only at the end, when the leading orders of the
expansion have been found, $\gamma$ is given its true value
$\gamma=\sqrt{\tau_m/\tau_s}$. 

The solution at zero-th order of the FPE (\ref{eq:FPE}) satisfying
conditions (\ref{eq:J},\ref{eq:FPE-z-sol}) is

\begin{equation}
 P_0(x,z,t|\hat H,z_0) = P(z,t|z_0) \delta(x-X(z,t))  \;,
   \label{eq:P_0}
\end{equation}

\noindent
where $X(z,t)$ is the time evolution of the variable $x$ obtained from
eq. (\ref{eq:x}) with {\em frozen} $z$ and initial condition $\hat
H$. Notice that $x=X(z,t)$ is a periodic function of $t$, because
whenever $x=\hat \Theta$, $x$ is reset to $\hat H$. Its period,
$T(z)=\tau_m ln(\hat H -\gamma z/\hat \Theta -\gamma z)$
($T(z)=\infty$ for $z<\hat \Theta/\gamma$), is the
inter-spike-interval (ISI) of a LIF neuron receiving a frozen $z$, and
it is calculated from eq. (\ref{eq:x}) as the first time $T$ at which
$X(z,T)=\hat \Theta$. After expressing the delta functions in terms of
$t$, the probability density current, eq. (\ref{eq:J}), at zero-th
order becomes

\begin{equation}
   J_0(z,t|z_0) = P(z,t|z_0) \sum_{n=1}^{\infty} \delta(t-nT(z))
   \label{eq:J_z_0} \;.
\end{equation}

\noindent
This expression has a simple interpretation. The sum of delta
functions in the index $n$ represents a regular train of spikes with
ISI $T(z)$, as if $z$ were fixed. Therefore, the probability of having a
spike along with a fluctuation $z$ at time $t$, $J_0$, is given at a
first approximation by the product of both the probability of finding
at time $t$ a spike of the train generated with frozen fluctuation
$z$, and the probability of having such a fluctuation $z$ at time $t$
starting from the initial condition $z=z_0$, $P(z,t|z_0)$. 
Note that in eq.(\ref{eq:J_z_0}) the noise is allowed to evolve in time 
following the distribution $P(z,t|z_0)$. It has been proved that 
the stationary (frozen) distribution of $z$ can be employed
to describe the firing rate of LIF neurons \cite{Mor+04,Mor+05}, 
and used the approximation that $z$ is constant during the ISIs 
to describe the Fano factor of non-LIF neurons with weak noise \cite{Mid+03}.
However, freezing completely the noise $z$ in eq.(\ref{eq:J_z_0}) 
leads to very poor predictions in our problem (not shown).

To determine the ACF, eq. (\ref{eq:C}), at zero-th order, we have to
the zero-th order $J(z)$ is required, which is \cite{Mor+04}

\begin{equation}
 J_0(z)= \nu_0(z) p(z) \;,
   \label{eq:J_0}
\end{equation}

\noindent
where $\nu_0(z)=1/T(z)$ for $z \ge \hat \Theta/\gamma$ and
$\nu_0(z)=0$ otherwise.  Finally, the zero-th
order ACF is computed, after using
eqs. (\ref{eq:C},\ref{eq:J_z_0},\ref{eq:J_0}) and evaluating the delta
functions, as

\begin{equation}
   C_0(t) = \sum_{n=1}^{\infty}  \int  
             \frac{ dz_0 \; J_0(z_0) \; 
                      (\gamma z_n-\hat H )(\gamma z_n-\hat \Theta )}
            { \nu_0 \tau_m n  \; \gamma \; (\hat \Theta - \hat H )}
           P(z_n,t|z_0)                
  \label{eq:C_0}
  \;,  
\end{equation}

\noindent
where $z_n \equiv z_n(t) \equiv \gamma^{-1}(\hat \Theta -\hat H
e^{-t/n\tau_m}) /(1-e^{-t/n\tau_m})$. The $z_n$s are the roots of the
equations $t=nT(z_n)$, the zeros of the delta functions in
eq. (\ref{eq:J_z_0}).

In Fig. (\ref{fig:grafica1}) we plot the ACF for the output spike train
of a LIF neuron computed using eq.(\ref{eq:C_0}) and compare it with
simulation results. The agreement is very good in both the
subthreshold (left) and suprathreshold (right) regimes. In both
regimes, the ACF shows a prominent peak after a relative refractory
period of about $10ms$ ($\sim\tau_m$). This means that
the potential has to be integrated from reset to threshold to emit the
first spike. The prominent peak indicates that the neuron is bursty,
producing spikes that are grouped within short time intervals of
$20ms$ ($\sim\tau_s$) \cite{Mor+04}. After the prominent peak, the
ACF decays to a steady-state value either
monotonically (left) or with a damped oscillation (right). Damped
oscillations are a robust feature in the suprathreshold regime, as is
their absence in the subthreshold regime. This reflects the fact that
the neuron in the suprathreshold regime fires more regularly, and
therefore the output spikes tend to occur at
integer number of times the mean ISI (see the peaks of the
oscillations in the ACF). For long times ($t \gg \tau_s$) the memory of the spike at time
$t=0$ has disappeared, and the ACF decays to the unconditioned probability of having a
spike, that is, the firing rate of the LIF neuron.

%%%%%%%%%%%%%%%%%%%%%%%%%%%%%

\begin{figure}
\includegraphics[width=8cm,height=5cm,angle=0]{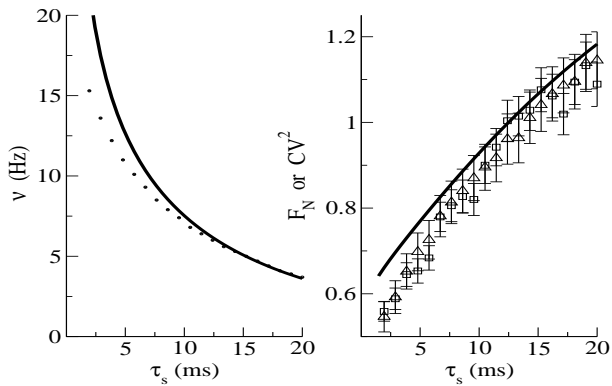}
\caption{\label{fig:grafica2}
The firing rate (left), the Fano factor (right), $F_N$, and the $CV^2$
(right) for the output spike train of a LIF neuron are plotted as a
function of $\tau_s$. The firing rate prediction (line) is calculated
using both eqs. (\ref{eq:C_0_lim}) and $\nu_0=\int dz J_0(z)$,
and it is compared with simulation results (points). The $F_N$ predicted by
eqs. (\ref{eq:F_N},\ref{eq:C_0},\ref{eq:C_0_lim}) (line) is compared
with the $F_N$ (squares) and $CV^2$ (triangles) obtained from
simulations. Parameters values are as in the subthreshold regime of
Fig. (\ref{fig:grafica1}).}
\end{figure}

\noindent
{\bf The firing rate, Fano factor and CV.}  As it is clear, the
firing rate can be obtained from the ACF, eq. (\ref{eq:C_0}), in the
limit of long times ($t \gg \tau_s$). This rate has the expression

\begin{equation}
   \nu_0 = \lim_{t \rightarrow \infty} \sum_{n=1}^{\infty} 
                 \frac{(\gamma z_n(t)-\hat H )(\gamma z_n(t)-\hat \Theta ) }
		{\sqrt{2 \pi}  \tau_m \;n \; \gamma (\hat \Theta - \hat H)} 
               \;
               e^{-z_n(t)^2/2}                        
  \label{eq:C_0_lim}
  \;.  
\end{equation}

\noindent
A different expression for the firing rate can be computed using
$\nu_0=\int dz J_0(z)$ \cite{Mor+04}. In fact, both expressions give
identical results when they are plotted as a function of $\tau_s$
(continuous curve in Fig. (\ref{fig:grafica2}), left). However,
computationally, eq. (\ref{eq:C_0_lim}) is much faster because it only
involves a sum that can be cut at $n \sim 200$ (using
$t=200ms$). Naturally, the number of terms needed to approximate the
ACF and the firing rate grows as $t$ increases. Comparison of both
expressions of $\nu_0$ with simulation results shows that the
prediction is very good even when $\tau_s \sim
\tau_m$.

The Fano factor of the output spike train, $F_N$, defined as the ratio
between the variance of the spike count and its mean evaluated for
long time windows, is directly related to the time integral of ACF as
(\cite{Cox62} and see, e.g., eq. (3) of ref \cite{Mor+02})

\begin{equation}
   F_{N} = 1 + 2 \int_{0}^{\infty} dt \; (C(t)-\nu )                 
  \label{eq:F_N}
  \;.  
\end{equation}

\noindent
We have evaluated the zero-th order $F_N$ in eq. (\ref{eq:F_N}) using
the zero-th order solutions of $C(t)$ and $\nu$,
eqs.(\ref{eq:C_0},\ref{eq:C_0_lim}). The prediction fits very well the
simulation results (right panel of Fig. (\ref{fig:grafica2})). We have
also computed the coefficient of variation of the ISIs, $CV$, of the
neuron response using simulations (same panel). It is known that for
renewal processes $F_N \equiv CV^2$ (e.g. for a Poisson
process $F_N \equiv CV^2=1$, and $C(t)=\nu$). Here we find that $F_N
\sim CV^2$ even when the output response is not a renewal
process. This is because, although the synaptic time scale introduces
correlations in the successive ISIs, since for low (but typical)
rates $\tau_s < \nu^{-1}$, the correlation between successive ISIs is
small. This explains the similarity between $F_N$ and $CV^2$. Notice
that the firing variability is large when $\tau_s \ge \tau_m$ \cite{Mor+05,Svi-Des}.

%%%%%%%%%%%%%%%%%%%%%%

\begin{figure}
\includegraphics[width=8cm,height=5cm,angle=0]{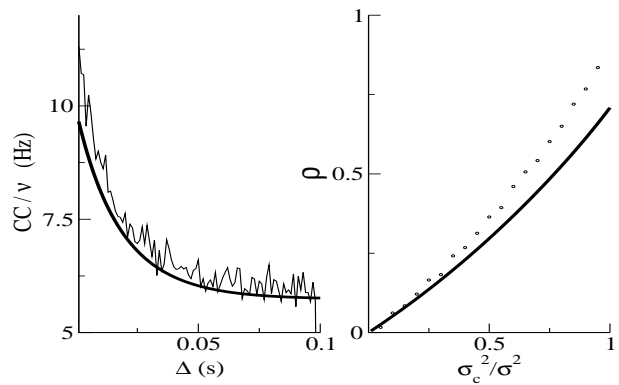}
\caption{\label{fig:grafica3}
{\bf Left:} Theoretical (thick line, eq.(\ref{eq:CCG_0})) and
simulated (thin line) CCFs normalized by the firing rate of one of
the neurons as a function of time lag. Here $\sigma_c^2=2Hz$. {\bf
Right:} Theoretical (line, eq. (\ref{eq:rho})) and simulated (points)
correlation coefficient, $\rho$, for the output spike trains of a pair
of identical LIF neurons as a function of the fraction of common
noise, $\sigma_c^2/\sigma^2$. The numerical $\rho$ is calculated using
eq. (\ref{eq:rho}) integrating the simulated CCF. Parameters for both
figures are $\mu=85Hz$, $\sigma^2=9Hz$, and the others as in
Fig. (\ref{fig:grafica1}).  }
\end{figure}

\noindent
{\bf The crosscorrelation function and correlation coefficient.} A
central issue to describe population dynamics is to understand the way
neuron activity synchronizes. Here we study a {\em pair of
identical LIF neurons} $(k=1,2)$
     
\begin{equation}
    \tau_m \; \dot{V_k}= -V_k + \tau_m \;( \; I_k(t)\;+\;I_c(t) \; )  \;, 
    \label{eq:V-2neu}
\end{equation}

\noindent

\noindent
receiving both an independent source of current, $I_k(t)$, and a
common source, $I_c(t)$. Each current is described by an equation
identical to eq. (\ref{eq:I}), with mean $\mu_{ind}$ and variance
$\sigma^2_{ind}$ for the independent components, and mean $\mu_c$ and
variance $\sigma^2_c$ for the common component. Each neuron receives a
total mean current $\mu=\mu_{ind}+\mu_c$ and total variance
$\sigma^2=\sigma^2_{ind}+\sigma^2_c$.

The CCF of the output spike trains of the two neurons (denoted as
$CC(\Delta)$) can be obtained by an analysis similar to that used for
the ACF. The CCF is defined as the joint probability density of having
a spike of neuron $1$ at a given time and a spike from neuron $2$
after a delay $\Delta$. Here we only summarize the main
results. First, we define the normalized fluctuations
$u_k=(I_k+I_c-\mu)/\sigma$, having zero mean and unit variance. 
Notice that these are not independent because of the common input 
$I_c$. Second, if neuron $1$ has a fluctuation $u_1$, the probability
density that after a delay $\Delta$ neuron $2$ has a fluctuation
$u_2$, $P(u_2,\Delta|u_1)$, is a Gaussian distribution with mean
$\left< u_2(\Delta,u_1) \right> =u_1 e^{-\Delta/\tau_s}
\sigma_c^2/\sigma^2$ and variance
$Var(u_2(\Delta))=1-e^{-2\Delta/\tau_s}\sigma^2_c/\sigma^2$. Then,
for long $\tau_s$

\begin{eqnarray}
    CC_0(\Delta)  =  \lim_{t \rightarrow \infty} 
          \int du_1 du_2 \; P(u_2,\Delta|u_1) \; p(u_1)  &&
\nonumber
\\ 
    \;\;\;\;\;\;\
     \sum_{n,m=1}^{\infty} \delta(t-nT(u_1)) \; \delta(t+\Delta-mT(u_2))
   \;,   &&
  \label{eq:CCG_u}  
\end{eqnarray}

\noindent
where $T(u_k)=\tau_m ln(\hat H -\gamma u_k/\hat \Theta -\gamma u_k)$
($T(u_k)=\infty$ for $u_k<\hat \Theta/\gamma$) is the ISI of the
neuron $i$ receiving a {\em constant} fluctuation $u_k$, and $p(u_1)$
is a normal distribution describing the steady state distribution of
the fluctuations of neuron $1$. The quantities $\gamma$, $\hat H$ and
$\hat \Theta$ are defined as before. The two sums of delta functions
in eq. (\ref{eq:CCG_u}) can be interpreted as the product of two
output spike trains with fixed ISI (determined by the input
fluctuations), quantity which has to be averaged over all the possible
fluctuations. The result of such an average is the CCF when the limit
$t \rightarrow \infty$ is taken to allow randomization of the initial
conditions, eq.(\ref{eq:CCG_u}). This equation can be simplified by
integration of the delta functions, obtaining

\begin{eqnarray}
   CC_0(\Delta)  = \lim_{t \rightarrow \infty}
       \sum_{n,m=1}^{\infty} 
           \frac{ (\gamma a_n-\hat H )(\gamma b_n-\hat H )}
              { n \; m\; \tau_m^2  \; \gamma^2 \; (\hat \Theta - \hat H )^2} 
\nonumber
\\     
   \;\;\;\;\; (\gamma a_m-\hat \Theta )(\gamma b_m-\hat \Theta )    
            P(b_m,\Delta|a_n) \; p(a_n)
  \label{eq:CCG_0}
  \;,  
\end{eqnarray}

\noindent
where $a_n \equiv z_n(t)$ and $b_m \equiv z_m(t+\Delta)$, with $z_n(t)$
as in eq. (\ref{eq:C_0}). The theoretical CCF matches very well the
simulated one (Fig. (\ref{fig:grafica3}), left). Typically, the
prediction underestimates the central peak occurring at time lag
zero. The central peak of the CCF decays within a time of the order of
$\tau_s$ (notice that the CCF is symmetric around $\Delta=0$). This is
because the synaptic input, being slower than the neuron dynamics,
sets its own time scale in the dynamics of interactions of the two
neurons. The existence of a single peak is robust for low values of
$\sigma_c^2$ in both the sub and suprathreshold regimes, but other
side secondary peaks arise when all the noise is essentially
common. For long $\Delta$, the CCF converges to the product of the
firing rates at zero-th order, $\nu_0^2$ (see eq. (\ref{eq:C_0_lim})),
because the neurons fire independently.

The correlation coefficient, $\rho$, of the spike counts for long time
windows of the output spike trains of two identical neurons can be
computed from their CCF (\cite{Cox62} and see, e.g., eq. (4) of ref
\cite{Mor+02})

\begin{eqnarray}
    \rho =  \frac{2} {F_N \nu} 
             \int_{0}^{\infty} ds \; (CC(s) - \nu^2)
    \label{eq:rho}
  \;.  
\end{eqnarray}

\noindent
For the two neurons in eq. (\ref{eq:V-2neu}) it can be
computed at zero-th order using the zero-th orders of $CC(\Delta)$,
eq.(\ref{eq:CCG_0}), $F_N$ and $\nu$. We have compared the theoretical
and simulated $\rho$ as the fraction of common noise increases
(Fig. (\ref{fig:grafica3}), right). The prediction is good for low
values of common noise, and departs from the simulations for larger
values. As the common noise increases, $\rho$ increases monotonically
and reaches $\rho=1$ when the common noise equals the total input
noise. Correlation coefficients of $\sim 0.1$ as those found in cortex
\cite{Sha+98otros} are predicted accurately, and they are obtained
when the common noise represents $\sim 20$ per cent of the total
synaptic noise entering into the neuron, which can be a realistic
value \cite{Sha+98otros}. Therefore, the right plot at Fig.3 provides
a valuable tool to estimate the fraction of common noise from the
correlations of the spike trains of pairs of neurons, a quantity which
otherwise is not available experimentally.

The results we have obtained at the cell level open the way for
a systematic investigation of the role of correlations in neuronal networks.

\indent 
R.M. thanks N. Rubin and J. Rinzel for their hospitality at
the CNS. We thank A. Renart for his comments. R.M. and N.P. are
supported by the Swartz Foundation and N.P. also by 
the Spanish grant BMF 2003-06242.

%%%%%%%  BIBLIOGRAFIA

\end{document}